\documentclass[letterpaper]{article} 
\usepackage{aaai25}  
\usepackage{times}  
\usepackage{helvet}  
\usepackage{courier}  
\usepackage[hyphens]{url}  
\usepackage{graphicx} 
\usepackage{natbib}  
\usepackage{caption} 
\usepackage{algorithm}
\usepackage{algorithmic}

\usepackage{newfloat}
\usepackage{listings}

\usepackage{mathrsfs}
\usepackage{amssymb}
\usepackage{bm}
\usepackage{amsmath}
\usepackage{dsfont}
\usepackage{booktabs}
\usepackage{epstopdf}
\usepackage{multirow}
\usepackage{comment}
\usepackage[table,xcdraw]{xcolor}
\usepackage{cite}
\usepackage{endnotes}
\usepackage{makecell}
\usepackage{subcaption}
\usepackage{adjustbox}
\DeclareCaptionStyle{ruled}{labelfont=normalfont,labelsep=colon,strut=off} 
\floatstyle{ruled}
\newfloat{listing}{tb}{lst}{}
\floatname{listing}{Listing}
\title{Knowledge Graph Completion with Relation-Aware Anchor Enhancement}
\author{
    Duanyang Yuan\textsuperscript{\rm 1}\equalcontrib, Sihang Zhou\textsuperscript{\rm 1}\equalcontrib, Xiaoshu Chen\textsuperscript{\rm 2}, Dong Wang\textsuperscript{\rm 1}, Ke Liang\textsuperscript{\rm 2}, \\
    Xinwang Liu\textsuperscript{\rm 2}, Jian Huang\textsuperscript{\rm 1}\thanks{Corresponding author.}\\
}
\affiliations{
    \textsuperscript{\rm 1}College of Intelligence Science and Technology,
    National University of Defense Technology, Changsha, China\\
    \textsuperscript{\rm 2}College of Computer Science and Technology, National University of Defense Technology, Changsha, China\\ 

    \{ydy\_n1, shzhou, chenxs, lizexin, liangke22, xinwangliu, huang\_jian\}@nudt.edu.cn
}

\usepackage{bibentry}

\begin{document}

\maketitle

\begin{abstract}
Text-based knowledge graph completion methods take advantage of pre-trained language models (PLM) to enhance intrinsic semantic connections of raw triplets with detailed text descriptions. Typical methods in this branch map an input query (textual descriptions associated with an entity and a relation) and its candidate entities into feature vectors, respectively, and then maximize the probability of valid triples. These methods are gaining promising performance and increasing attention for the rapid development of large language models. According to the property of the language models, the more related and specific context information the input query provides, the more discriminative the resultant embedding will be. In this paper, through observation and validation, we find a neglected fact that the relation-aware neighbors of the head entities in queries could act as effective contexts for more precise link prediction. Driven by this finding, we propose a relation-aware anchor enhanced knowledge graph completion method (RAA-KGC). Specifically, in our method, to provide a reference of what might the target entity be like, we first generate anchor entities within the relation-aware neighborhood of the head entity. Then, by pulling the query embedding towards the neighborhoods of the anchors, it is tuned to be more discriminative for target entity matching. The results of our extensive experiments not only validate the efficacy of RAA-KGC but also reveal that by integrating our relation-aware anchor enhancement strategy, the performance of current leading methods can be notably enhanced without substantial modifications. 
\end{abstract}
\begin{links}
\link{Code}{https://github.com/DayanaYuan/RAA-KGC}
\end{links}

\section{Introduction}

A knowledge graph is a structured semantic knowledge base that stores world knowledge through triples, i.e., ( \textit{head entity},  \textit{relation},  \textit{tail entity}), or (\textit{h}, \textit{r}, \textit{t}) for short \cite{fan2024flow}. It is the cornerstone of many important applications \cite{liang2023knowledge}, such as question answering \cite{hao2017end, xu2022text}, information retrieval \cite{qin2023towards, liu2018entity} and recommendation system \cite{zhang2016collaborative, chen2023temporal}. Due to the intricate relations between entities, manually constructed knowledge graphs often struggle to encompass knowledge comprehensively. As a consequence, knowledge graph completion, which uncovers missing links by reasoning over known facts, is an important task for enhancing the application effects in this field.

\begin{figure} 
\centering
\includegraphics{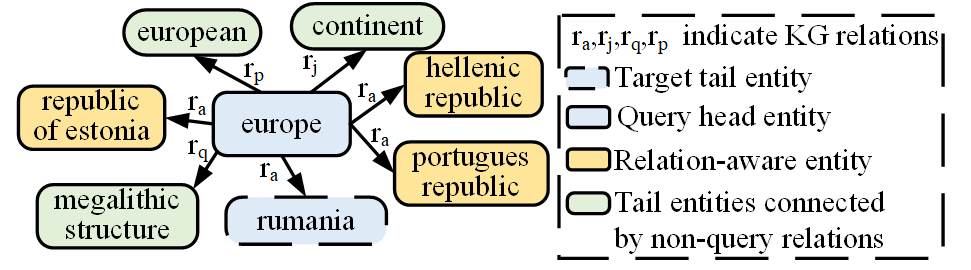}
\caption{In this figure, the yellow bricks denote relation-aware entities which share the same head entity and relation in the testing triple. These entities share important concept information with the target entity (the dashed blue brick), thus can be used to guide for better link prediction.}
\label{1induction_triple}
\end{figure}

The existing methods in this field can be broadly divided into two categories, i.e., triple-based \cite{li2024triplet,wang2014knowledge} and text-based methods \cite{wang-etal-2022-simkgc, shang2023relation}. Algorithms in the second category, which utilize pre-trained language models to incorporate additional relations and entity descriptions for improved predictive accuracy, have garnered escalating interest in recent years. Specifically, the text embedding methods \cite{wang-etal-2022-simkgc,yao2019kg}, as typical representatives of text-based methods, are achieving promising performance with appealing efficiency by adopting a bi-encoder network structure and a contrastive learning manner.
In these methods, the visible components of the triplets, either (\textit{h}, \textit{r}, ?) or (?, \textit{r}, \textit{t}), together with the corresponding descriptions serve as queries, while the tail or head entities considered as the candidate values. Employing language models, this branch's methods map queries and candidate values into a shared hidden space with structure identical but independent sub-networks. They perform link prediction by aligning the embeddings of valid triples and repelling those of the invalid ones. Our proposed method also falls into this group.

The performance of these methods is highly related to two vital factors. First, the representative capability of the pre-trained language model. To this aspect, thanks to the rapid development of large language models \cite{devlin-etal-2019-bert, openai2024gpt4technicalreport}, their stunning knowledge storage and context understanding ability have ensured the representative capability of the learned embeddings. Second, the precision of the query. While finding the appropriate value to match the query, it is easy to know that the more precise context information that the query provides, the easier the matching will be. Take the tail prediction task as an example. According to the setting of the task, the included descriptions of head entities and relations have largely specified these two components. However, to the tail entity, not even a word of the what the target entity looks like is provided. To solve the problem, the existing methods design various negative sampling algorithms to enable the learned network to better tell what does the valid tail entities not look like \cite{jiang2023don, li2023kermit, he2023mocosa, je2023unifying}. Although good performance has been acquired, the existing methods neglect to provide context information of the tail entities.

In this paper, we find that the relation-aware entities can act as effective contexts of the target entities. In Figure. \ref{1induction_triple}, we illustrate a sub-graph of the WN18RR dataset as an example. In this subgraph, \textit{rumania} is the target tail entity of the incomplete triplet (\textit{europe}, \textit{has part}, ?) in the testing set. The yellow entities represent the neighbors of the query head entity that share the same relation with the target tail entity. These entities (connected with the query head entity by the query relation) are denoted as relation-aware entities in our paper. Through this figure, we have two observations. 
1) Relation-aware entities are more inclined to possess similar semantic concepts, introducing these entities as contexts could provide important guidance to the target entity prediction. For example, in this subgraph, when given the relation-aware entity \textit{portugues republic}, it is more intuitive to anticipate the algorithm providing a European country as a response to the query (\textit{europe}, \textit{has part}, ?). 
These entities can help specify the query to make them to be more different from similar counterparts, like (\textit{continent}, \textit{has part}, ?). 
2) Without introduction of the target entity, it is very likely to have multiple entities that is suitable for a query. For example, besides Rumania, Germany, and Spain are all suitable for the query. In this circumstance, a possible solution to improve the algorithm performance is to enhance the average matching degree of the whole relation-aware entity set.

Driven by the observations, we randomly sample entities from the relation-aware entity set as anchors to guide the query embedding learning. Specifically, in our method, we also adopt a bi-encoder network structure and a contrastive learning manner \cite{wang-etal-2022-simkgc}. 
However, besides constructing classic queries which only contain information of head entities and relations, more context abundant anchor-enhanced queries are developed. These queries are formulated by appending anchor entities and their descriptions to the original query. With these queries, the network is trained by minimizing the InfoNCE loss \cite{chen2020simple} over both classic and anchor-enhanced query embeddings. In this setting, with the anchor information as extra context of the tail entity, the generated anchor-enhanced embedding is pulled to get closer to the neighborhood of the target entity, thus being more discriminative over the link prediction task. Moreover, as the relation-aware entities are included in the query sequence, more relation compact representation will be acquired. The contributions of this paper are twofold: 
\begin{itemize}
    \item Through observation and experimental verification, we find a previously overlooked fact that the relation-aware entities can act as effective contexts of target entity prediction. Enhanced by this setting, the current leading methods can achieve significant performance improvements without requiring substantial modifications.
    \item We propose a relation-aware anchor enhanced knowledge graph completion algorithm (RAA-KGC) by complementing the queries with target entity prototypes. The proposed method is proven to generate discriminative and compact embeddings, thus achieves superior performance against the compared state-of-the-art algorithms.
\end{itemize}

\section{Related Work}
\subsection{Knowledge Graph Completion}
The KGC methodology has experienced rapid advancement in recent years. To the triple-based KGC methods, numerous strategies (i.e., translation-based \cite{bordes2013translating, wang2014knowledge}, semantic matching-based \cite{Nickel2011ICML}, neural network-based \cite{schlichtkrull2018modeling, li2023event, fan2024relation}, logical-rule based \cite{sadeghian2019drum}) have been developed to construct suitable entity and relation embeddings while conducting proper assessment to the validation of the candidate triples. 
These models acquire the structural aspects of a knowledge graph, excluding the textual content of entities and relations. Comparatively, text-based methods suggest integrating textual knowledge of entities and relations into the model via Pre-trained Language Models \cite{zeng2024kbada}.
The pioneer text-based method, DKRL \cite{xie2016representation}, initially integrates textual descriptions into entity embeddings with a convolutional neural network. Subsequent studies, including KG-BERT \cite{yao2019kg} and KEPLER \cite{wang2021kepler}, utilize a Pre-trained Language Model (PLM) for encoding text descriptions. Moreover, KGT5-context \cite{kochsiek2023friendly} and KG-S2S \cite{chen-etal-2022-knowledge} model the link prediction task as a sequence-to-sequence generation problem and output the entity predictions by the decoder directly. Furthermore, LMKE \cite{ijcai2022p318} introduces a contrastive learning framework that employs a PLM to derive entity and relation embeddings within the same space as word tokens, showcasing its efficacy in addressing the long-tail issue. Good performance have been achieved by the text-based algorithms, and the performance is highly correlated with the context information which is given to depict the target entities.

\subsection{Context Information Adopted by KGC}
Precise context information is important for the rationality and accuracy of the output of the language models, that is why prompt engineering is so important to neural language processing \cite{li-etal-2023-robust}. To collect context information for the algorithms, a variety of methods have been proposed. The early methods straightforwardly take the descriptions of query entities and relations as context. After that, algorithms are proposed to collect more context information within the neighborhood of the query triplet. Among them, KGT5-context \cite{kochsiek2023friendly} takes the one-hop neighborhood of the query as extra context to the model. ConKGC \cite{shang2023relation} introduces a sampling strategy to extract weak positive samples within the relation-aware neighborhood of the query. BERTRL \cite{zha2022inductive} collects possible reasoning paths in the query neighborhood for intrinsic supervision. Besides the mentioned algorithms, GenKGC \cite{xie2022discrimination} and KICGPT \cite{wei2023kicgpt} go beyond the neighborhood of the query and take triples connected by the query relation as prediction demonstrations. Significant performance enhancement has been witnessed with the aforementioned context information extraction mechanism. However, these methods mainly focus on introducing more common knowledge, which is related to either the query entity or the query relation. In this paper, through our observation and experiments, we find that exploiting task-specific information besides the common knowledge information for the query is also important to the model performance. Integrating the proposed relation-aware anchor strategy with the existing mechanisms can further improve the algorithm performance. In the following section, we introduce the anchor generation procedure and the network structure of our method.

\section{Method}
\begin{figure*}[t] 
\centering
\includegraphics{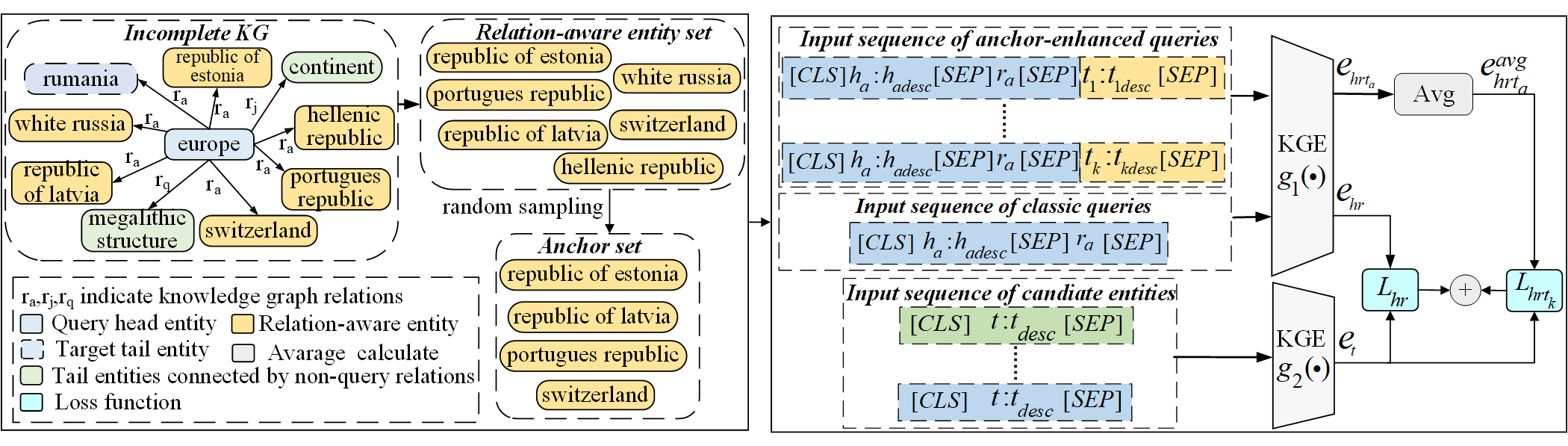}
\caption{Framework illustration of the proposed RAA-KGC. In this figure, the yellow bricks denote relation-aware entities that share the same head entity and relation. The dashed blue brick denotes the target entity. The core idea of RAA-KGC is to generate a general example of what might the tail (or head) entity be like by using relation-aware entities, and then pull its embedding towards the neighborhood of the anchor.}
\label{1method_overall}
\end{figure*}

\subsection{Problem Statement}
Given a knowledge graph (KG) $\mathcal{G}=\{(h,r,t)\big|h,t\in\mathcal{E},r\in\mathcal{R}\}$, where $\mathcal{E}$ and $\mathcal{R}$ are the set of entity and relation of the KG, respectively. \textit{h} and \textit{t} are the head and tail entities, while \textit{r} is the relation between them. The KGC tasks are usually divided into link prediction, triple classification, and relation prediction \cite{lin2015learning,yao2019kg, chen2022hybrid}. In this paper, we focus on the link prediction sub-task, which aims to predict the missing  \textit{h} in head entity prediction  (?,\textit{r}, \textit{t})  or \textit{t} in tail entity prediction (\textit{h}, \textit{r}, ?) for a relation fact triple (\textit{h}, \textit{r}, \textit{t}). The evaluation protocol is to rank all candidate entities instead of only giving one best result. Following the approach presented in SimKGC \cite{wang-etal-2022-simkgc}, we use the inverse triples (\textit{t}, $r^{-1}$, \textit{h}) to simplify the task by focusing exclusively on tail entity prediction, where $r^{-1}$ denotes the inverse relation of \textit{r}. Besides, we also summarize the notations in the Appendix.

\subsection{Anchor Generation}
The success of RAA-KGC is attributed to leveraging information from relation-aware entities, specifically utilizing the neighbors of the query head entity that share the same relation with the target tail entity. The RAA-KGC framework is illustrated in Figure. \ref{1method_overall}. Based on KGs, we define anchor generation operation as follows.

\textbf{Definition 1. Relation-Aware Anchor Generation} Given a knowledge graph $\mathcal{G}=\{(h,r,t)\big|h,t\in\mathcal{E},r\in\mathcal{R}\}$ and the corresponding inverse knowledge graph $\mathcal{G}_{inv}=\{(t,r^{-1},h)\Big|\forall(h,r,t)\in\mathcal{G}\}$. When the query is $(h_a,r_a,?)$, the relation-aware entity set $\mathcal{T}$ is the set of tail entities connected with the query head entity $h_a$ and relation $r_a$ on $\mathcal{G}\cup\mathcal{G}_{in\nu}$. The aim of anchor generation is to obtain anchor set $\mathcal{T}_k$, and the operation as follows.
\begin{equation}
\mathcal{T}_k=random(\mathcal{T},k),k\leq K, K=5
\end{equation}
where $\mathcal{T}_k=\{t_i|i=\{1,2,...,k\}, t_i\in\mathcal{T}\}$, $k$ is the random sampling size.

Given an input query (\textit{europe}, \textit{has part}, ?), the proposed method traverses all triples started from the query head entity $h_a$ in the given KG and only keeps the entities satisfying Def. 1. After that, we construct the anchor-enhanced queries $I_{hrt_a}^a$ by equation \eqref{hrt} and the classic query $I_{hr}^{a}$ by equation \eqref{hr}. We input these information into subsequent models.

\begin{equation}
\begin{split}
I_{hrt_a}^a = &\ [CLS]h_a:h_{adesc}[SEP]r_a[SEP] \\
              &\ t_i:t_{idesc}[SEP], i=\{1,2,\dots,k\}
\end{split}
\label{hrt}
\end{equation}

\begin{equation}
I_{hr}^{a}=[CLS]h_a:h_{adesc}[SEP]r_{a}[SEP]
\label{hr}
\end{equation}

\begin{equation}
I_{t}=[CLS]t:t_{desc}[SEP]
\end{equation}
where $h_a$ denotes query head entity, $t_i$ is entity in anchor set $\mathcal{T}_k$, $h_{adesc}$ and $t_{idesc}$ are the descriptions of $h_a$ and $t_i$, $r_a$ is the relation between $h_a$ and $t_i$, $t$ and $t_{desc}$ are candiate entities and their descriptions.

\subsection{Anchor-Enhanced Embedding Generation}

Following the current state-of-the-art KGC methods \cite{wang-etal-2022-simkgc}, we use two encoders, $g_{_1}(\cdot)$ and $g_{_2}(\cdot)$, both initialized with the bert-base-uncased model but do not share parameters. The anchor-enhanced embedding is computed by Def.2. In addition, the $I_{hr}^{a}$ is encoded by encoder $g_{_1}(\cdot)$ to generate a relation-aware embedding $\bm{e_{hr}}$. The second encoder $g_{_2}(\cdot)$ is used to compute candidate embedding $\bm{e_t}$ for candidate entities sequence $I_{t}$. 

\textbf{Definition 2. Anchor-Enhanced Embedding Generation} Given a knowledge graph $\mathcal{G}=\{(h,r,t)\big|h,t\in\mathcal{E},r\in\mathcal{R}\}$ and the corresponding inverse knowledge graph $\mathcal{G}_{inv}=\{(t,r^{-1},h)\Big|\forall(h,r,t)\in\mathcal{G}\}$. 
$I_{hrt_a}^a$ is encoded by encorder $g_{_1}(\cdot)$  as $\bm{e_{hrt_{a}}}$. Then, the operation of generating anchor-enhanced embedding $\bm{e_{hrt_{a}}^{avg}}$ as follows.
\begin{equation}
\bm{e_{hrt_{a}}^{avg}}=average(\bm{e_{hrt_{a}}})
\end{equation}
where $\bm{e_{hrt_{a}}}$ is anchor-enhanced queries embedding. 

\subsection{Training and Inference}

We formulate a simple yet effective contrastive learning loss function based on the anchor-enhanced embedding, which uses cosine to measure the similarity between the target entity and the query \cite{9385922}. The function leverages the hidden semantics underlying the relation-aware entities, as shown in equations \eqref{eql1}-\eqref{eql3}. The positive samples are the fact triples in KG, and the negative samples are the self-negative samples (SN) and in-batch negative samples (IBN) \cite{wang-etal-2022-simkgc}. Moreover, different relations within the same batch will have different anchor sets, leading to different anchor-enhanced embedding, even if the query head entity is the same. Therefore, we also incorporate in-batch-relation negative samples (IBRN). This IBRN method is similar to IBN. By minimizing $L_{hrt_{a}}$, the anchor-enhanced embedding $\bm{e_{hrt_{a}}^{avg}}$ can be further rotated and scaled, thereby bringing the target entity embedding closer to $\bm{e_{hrt_{a}}^{avg}}$ in the resulting semantic space. Based on this, when predicting $(h,r,?)$, $\bm{e_{hrt_{a}}^{avg}}$ and target entity embedding will get a higher score than others, so we can prioritize search results in these relation-aware entities. After the above operation, the generated anchor-enhanced embedding is pulled toward the neighborhood of the target entity. Then, it is positioned to the specific anchor through $L_{hr}$.

\begin{equation}
L_{cls}=\alpha{\cdot}L_{hrt_{a}}+L_{hr}
\label{eql1}
\end{equation}

\begin{equation}
L_{hrt_{a}}=-\log\frac{e^{(\phi(\mathrm{h},\mathrm{r},\mathrm{t}_{a})-\gamma)/\tau}}{e^{(\phi(\mathrm{h},\mathrm{r},\mathrm{t}_{a})-\gamma)/\tau}+\sum_{t_{a}^{\prime}\in\varphi}e^{\phi(\mathrm{h},\mathrm{r},\mathrm{t}_{a}^{\prime})/\tau}}
\label{eql2}
\end{equation}

\begin{equation}
L_{_{hr}}=-\log\frac{e^{(\phi(h,r,t)-\gamma)/\tau}}{e^{(\phi(h,r,t)-\gamma)/\tau}+\sum_{t^{\prime}\in\delta}e^{\phi(h,r,t^{\prime})/\tau}}
\label{eql3}
\end{equation}
where $\varphi=(\mathrm{IBN}\cup\mathrm{IBRN})$, $\delta=(\mathrm{SN\cup IBN})$, $\gamma$ is the additive margin, $\tau$ is a learnable temperature parameter, $\alpha$ represents the trade-off weight, $\phi(h,r,t_a)$ is cosine of $\bm{e_{hrt_{a}}^{avg}}$ and $\bm{e_t}$, $\phi(h,r,t)$ is cosine of $\bm{e_{hr}}$ and $\bm{e_t}$.

For inference, we calculate the two cosine similarity, $\phi(h,r,t_a)$ and $\phi(h,r,t)$, and choose the sum of the highest score of each as the prediction: 

\begin{equation}
\text{argmax }\phi(h,r,t_a)+\phi(h,r,t)
\end{equation}

\section{Experiments}

This section is organized as follows. Firstly, we introduce the experimental settings. Then, we evaluate RAA-KGC’s performance on three benchmark datasets. Finally, we conduct some experimental studies to investigate: (i) the importance of anchor-enhanced embedding, (ii) relation-wised performance, (iii) the impact of sample size, (iv) the compatibility of the relation-aware anchor enhancement mechanism, (v) performance over unseen entities. (vi) case study.

\subsection{Experimental Settings}
\subsubsection{Datasets}

We evaluate RAA-KGC on three commonly used datasets: WN18RR \cite{dettmers2018convolutional}, FB15k-237 \cite{toutanova2015observed}, Wikidata5M-Trans \cite{wang2021kepler}. The details of these three datasets are shown in Table \ref{DATASET_INFO1}. For WN18RR and FB15k-237 datasets, Teru et al. \cite{teru2020inductive} extracted four versions of transductive and inductive datasets, respectively. For WN18RR and FB15k-237 datasets, we utilize the text descriptions provided by KG-BERT \cite{yao2019kg}. The Wikidata5M-Trans dataset already contains descriptions for all entities and relations.

For the WN18RR dataset, SimKGC \cite{wang-etal-2022-simkgc} preprocessed entities as follows. If the entity is $reform\_NN\_1$, it is extracted as $reform$. A large number of similar entities, such as $reform\_VB\_1$, $reform\_NN\_3$, and $reform\_VB\_5$, exist with distinct meanings. These account for approximately 13.76\% of the total entities. To preserve entity uniqueness, we do not apply this preprocessing.

\begin{table}
\centering{
\begin{tabular}{@{}cccccc@{}}
\hline
\textbf{Dataset} & \textbf{train} & \textbf{valid}  & \textbf{test} \\ \hline
\textbf{WN18RR} & 86,835       & 3034   & 3134       \\

\textbf{FB15k-237}  & 272,115       & 17,535  & 20,466       \\
\textbf{Wikidata5M-Trans}     & 20,614,279      & 5,163  & 5,163       \\\hline
\end{tabular}}
\caption{Statistics of the datasets used in this paper.}
\label{DATASET_INFO1} 
\end{table}

\subsubsection{Implementation Detail}

We employ the text-based model SimKGC\cite{wang-etal-2022-simkgc} as the baseline. The parameter settings for the KGE encoder are the same as shown in the original paper. We implement RAA-KGC based on the PyTorch library \cite{paszke2019pytorch}. The batch size is 32. As for the specific hyperparameters used in our work, we search the upper bound of the trade-off weight $\alpha$ for contrastive loss within the range \{0.1, 0.2, 0.3, 0.4, 0.5\}. We assess RAA-KGC using four automated evaluation metrics: Mean Reciprocal Rank (MRR) and Hit@k(k=1,3,10).

\subsubsection{Compared Baselines}

To demonstrate the superior performance of the proposed method, we compare it with twelve models. For triple-based methods, we reporte results of TransE \cite{bordes2013translating}, ComplEx \cite{trouillon2017complex}, DistMult \cite{yang2015embedding}, ConvE \cite{dettmers2018convolutional}, RGCN \cite{schlichtkrull2018modeling}, RotatE \cite{sunrotate}, and QuatE \cite{li2023knowledge}. For text-based methods, we include KG-BERT \cite{yao2019kg}, MTL-KGC \cite{kim2020multi}, StAR \cite{wang2021structure}, SimKGC \cite{wang-etal-2022-simkgc}, and HaSa \cite{zhang2024hasa}. Note that except for the results of SimKGC, the results of the compared baselines are recorded from the original papers.

\subsection{Link Prediction Performance}

\begin{table*}[ht]
\fontsize{9pt}{11pt}\selectfont
\begin{tabular}{ccccccccccccc}
\hline
\multirow{2}{*}{Methods} & \multicolumn{4}{c}{WN18RR}    & \multicolumn{4}{c}{FB15k-237} & \multicolumn{4}{c}{Wikidata5M-Trans} \\ \cline{2-13}
                         & MRR   & Hit@1   & Hit@3   & Hit@10  & MRR   & Hit@1   & Hit@3   & Hit@10  & MRR    & Hit@1   & Hit@3   & Hit@10  \\ \hline
\multicolumn{13}{c}{Triple-based   methods}                                                                            \\ \hline
TransE            & 24.3  & 4.3   & 44.1  & 53.2  & 27.9  & 19.8  & 37.6  & 44.1  & 25.3   & 17.0  & 31.1  & 39.2  \\
ComplEx         & 46.8  & 42.7  & 48.5  & 55.4  & 27.8  & 19.4  & 29.7  & 45.0  & 28.2   & 22.6  & -     & 39.7  \\
DistMult          & 44.4  & 41.2  & 47.0  & 50.4  & 28.1  & 19.9  & 30.1  & 44.6  & 33.9   & 27.1  & -     & 45.9  \\
ConvE             & 45.6  & 41.9  & 47.0  & 53.1  & 31.2  & 22.5  & 34.1  & 49.7  & -      & -     & -     & -     \\
RGCN            & 42.7  & 38.2  & 44.6  & 51.0  & 24.8  & 15.3  & 25.8  & 41.4  & -      & -     & -     & -     \\
RotatE            & 47.6  & 42.8  & 49.2  & 57.1  & \textbf{33.8}  & \textbf{24.1}  & \textbf{37.5}  & \textbf{53.3}  & 29.0   & 23.4  & 32.2  & 39.0  \\
QuatE            & 48.1  & 43.6  & -     & 56.4  & 31.1  & 22.1  & -     & 49.5  & -      & -     & -     & -     \\ \hline
\multicolumn{13}{c}{Text-based   methods}                                                                                 \\ \hline
KG-BERT           & 21.6  & 4.1   & 30.2  & 52.4  & -     & -     & -     & 42.0  & -      & -     & -     & -     \\
MTL-KGC           & 33.1  & 20.3  & 38.3  & 59.7  & 26.7  & 17.2  & 29.8  & 45.8  & -      & -     & -     & -     \\
StAR              & 40.1  & 24.3  & 49.1  & 70.9  & 29.6  & 20.5  & 32.2  & 48.2  & -      & -     & -     & -     \\
SimKGC            & 55.31 & 45.03 & 61.54 & 73.65 & 29.89 & 21.27 & 32.18 & 47.68 & 29.18  & 25.37 & 30.58 & 44.9  \\
HaSa            & 53.8  & 44.4  & 58.8  & 71.3  & 30.4  & 22.0  & 32.5  & 48.3  & -      & -     & -     & -     \\
RAA-KGC                & \underline{\textbf{59.74}}  & \underline{ \textbf{50.64}} & \underline{ \textbf{64.98}} & \underline{ \textbf{76.01}} & \underline{ 31.46} & \underline{22.38} & \underline{34.06} & \underline{49.26} & \underline{\textbf{34.15}}  & \underline{\textbf{29.24}} & \underline{ \textbf{34.47}} & \underline{\textbf{47.82}} \\ \hline
\end{tabular}
\caption{The results of WN18RR, FB15k-237, Wikidata5M-Trans datasets. Best performance is indicated by the bold face number in each column, and the underline means the best result compared to SimKGC in each column.}
\label{compasion_exper1}
\end{table*} 

To evaluate the effectiveness of the RAA-KGC, we compare it against twelve SOTA methods over three KGC datasets, including triple-based and text-based methods. The main results are summarized in Table \ref{compasion_exper1}, from which several conclusions can be drawn. First, RAA-KGC outperforms prior works on the WN18RR and Wikidata5M-Trans datasets. Specifically, on the WN18RR dataset, it improves MRR by 4.43\%, Hit@1 by 5.61\%, Hit@3 by 3.44\%, and Hit@10 by 2.36\% compared to SimKGC. Second, we observe that RAA-KGC falls short on the FB15k-237 dataset, possibly because the topological structure of this dataset is considerably denser \cite{yang2024enhancing, li2023kermit, cao2021missing}. Conversely, this kind of dataset is suitable for combining text and structure information for prediction \cite{he2023mocosa}, which we consider as a future research. Additionally, compared with QuatE in the triple-based methods, RAA-KGC improves MRR by 11.64\%, Hit@1 by 7.04\%, and Hit@10 by 19.64\% on the WN18RR dataset. Similarly, compared with TransE, RAA-KGC increases MRR by 35.44\%, Hit@1 by 46.34\%, and Hit@10 by 22.81\% on the WN18RR dataset. 
In the text-based methods, RAA-KGC achieves superior performance on all metrics of the WN18RR dataset and the Wikidata5M-Trans dataset. These results indicate entities similar to the target entity exist in the dataset, suggesting that leveraging these entities can enhance the model's ability to learn entity characteristics. 

\subsection{Importance of Anchor-Enhanced Embedding}

\begin{table}
\centering{
\begin{tabular}{ccccccccccccc}
\hline
\multirow{1}{*}{Metrics} & NT      & NRT     & IET  & IRT    \\ \hline
MRR                      & 55.31   & 50.99  & 51.66 & \textbf{59.74}  \\ 
Hit@1                    & 45.03   & 40.03  & 41.05 & \textbf{50.64}  \\ 
Hit@3                    & 61.54   & 57.21  & 60.47 & \textbf{64.98}  \\ 
Hit@10                   & 73.65   & 71.77  & 74.13 & \textbf{76.01}   \\ \hline
\end{tabular}
}
\caption{Results of adding different tail entity embeddings on the WN18RR dataset. The bold face number in each row indicates the best performance.}
\label{compasion_inrela}
\end{table}

\begin{figure}[H]
  \footnotesize
  \centering
  \begin{subfigure}[b]{.48\linewidth}
    \includegraphics{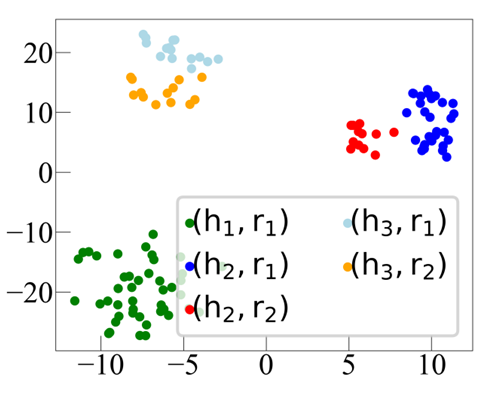}
    \caption{IRT}
    \label{fig:irt}
  \end{subfigure}%
  \begin{subfigure}[b]{.48\linewidth}
    \includegraphics{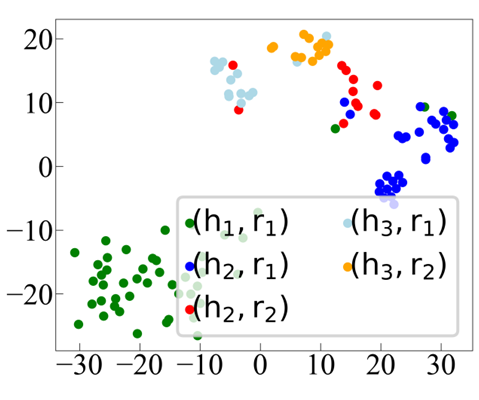}
    \caption{NT}
    \label{fig:nt}
  \end{subfigure}
  \caption{Visualization of entity embeddings with the different head entities and relations using t-SNE in NT and IRT scenarios. Points in same color indicate that their head entities and relations are the same. $h_1$, $h_2$, $h_3$ represent different query head entities, $r_1$ and $r_2$ indicate knowledge graph relations. The NMI of (a) and (b) are 0.75 and 0.62, respectively.}
  \label{1induction_tsne}
\end{figure}

In order to further validate the importance of anchor-enhanced embedding, the following experiments are conducted: (1) NT: uses only relation-aware embedding, (2) NRT: uses only tail entities connected by non-query relations, (3) IET: uses only anchor-enhanced embedding, (4) IRT: uses anchor-enhanced embedding and relation-aware embedding. Table \ref{compasion_inrela} shows these results. Besides, we select five groups of the tail entity sets from WN18RR dataset and embed them in both NT and IRT scenarios. The embeddings are visualized using t-SNE \cite{van2008visualizing}, as shown in Figure. \ref{1induction_tsne}. 

As shown in Table \ref{compasion_inrela}, the performance is optimal in the IRT scenario and minimal in the NRT scenario. These findings indicate that the relation-aware anchor plays an important role in the prediction process.
Meanwhile, we can see the following points from Figure. \ref{1induction_tsne}: (1) The gap among tail entities, which are connected by the same query head entity and different relations, is smaller than the gap among tail entities corresponding to different query head entities. For example, the gap between $(h_2, r_1)$ and $(h_2,r_2)$ is smaller than that between $(h_1, r_1)$ and $(h_2, r_1)$. Similar situations can be observed in the relations between other entities. (2) The comparison of IRT and NRT shows that the similarity between the neighbors of the query head entity and the target entity is higher than for others. Therefore, making full use of these entities allows the model to learn a general example of what might the tail (or head) entity be like and thus improves the model’s discrimination. (3) The superior performance of IRT compared to NT indicates that incorporating relation-aware entities offers the model additional supervision, thereby enhancing its performance.

\subsection{Relation-Wised Performance}

\begin{figure} 
\centering
\includegraphics{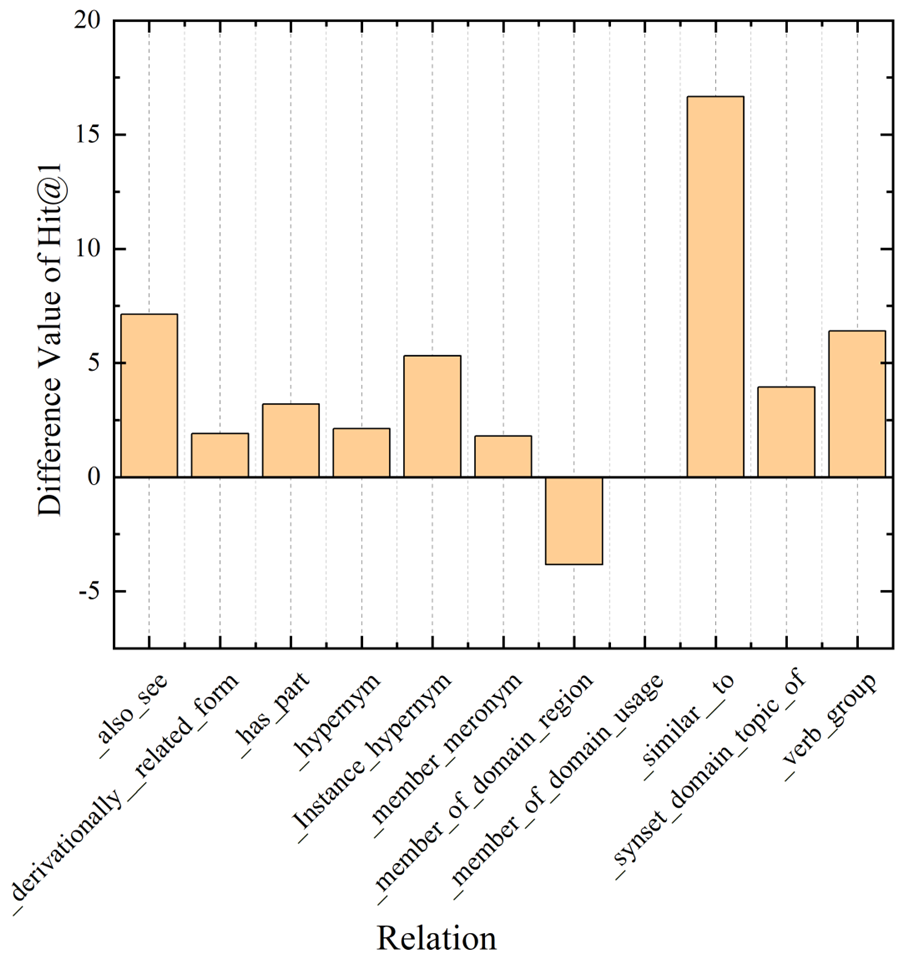}
\caption{The difference value of Hit@1 between RAA-KGC and SimKGC on different relations in the WN18RR dataset.}
\label{1relation_Category2}
\end{figure}

To explore which relation enhances the RAA-KGC performance from a fine-grained perspective, we compare the performance of the RAA-KGC and SimKGC for each relation on WN18RR dataset. The difference in Hit@1 between RAA-KGC and SimKGC is illustrated in Figure. \ref{1relation_Category2}. The details of MRR, Hit@1, Hit@3, and Hit@10 are provided in the Appendix.

As shown in Figure. \ref{1relation_Category2}, RAA-KGC outperforms SimKGC across most relation types, indicating the feasibility and effectiveness of our approach. Although it exhibits suboptimal performance on the $\_member\_of\_domain\_region$ relation, this relation types account for 1.06\% of all triples. Overall, these fine-grained analysis results show that RAA-KGC achieves outstanding performance.

\subsection{Impact of the Sample Size}

\begin{table}
\fontsize{9pt}{11pt}\selectfont
\begin{tabular}{ccccccc}
\hline
Metrics  & 0     & 1     & 2     & 3     & 4     & 5      \\ \hline
MRR                      & 55.31 & 55.69 & 55.75  & 58.08 & \textbf{59.74} & 59.56 \\ 
Hit@1                    & 45.03 & 45.36 & 46.19 & 48.63 & \textbf{50.64} & 50.48 \\ 
Hit@3                    & 61.54 & 62.76 & 63.99 & 63.78 & \textbf{64.98} & 64.93 \\ 
Hit@10                   & 73.65 & 73.87 & 74.46 & 75.19 & \textbf{76.01} & 75.62  \\ \hline
\end{tabular}
\caption{Impact of the sample size on WN18RR dataset. The bold face in each row indicates the best performance.}
\label{compasion_numbertail1}
\end{table}

We calculate the MRR and Hit@k(k=1,3,10) for different sample sizes on WN18RR datasets, as shown in Table \ref{compasion_numbertail1}. By analyzing these results, the performance is poorest at a sample size of 0, with performance improving as the sample size increases. When the sample size is 4, Hit@10 reaches 76.01\%, a 2.36\%  improvement over a sample size of 0 on WN18RR dataset. We attribute the improvement to the finding of anchor-enhanced embeddings, which capture more masked entities and contextual information for reconstructing the masked elements. However, as the sample size grows, the introduction of noise limits further performance gains. When the sample size is 5, Hit@10 is 75.62\% on WN18RR datasets, which is 0.39\% lower than the highest recorded values. The decline may be due to the presence of too many relation-aware entities, leading to a bias in the model's cognition of tail entities.

\subsection{Compatibility of the Relation-Aware Anchor Enhancement Mechanism}

\begin{table}
\begin{tabular}{ccccccccccccc}
\hline
\multirow{1}{*}{Methods} & MRR   & Hit@1 & Hit@3 & Hit@10   \\ \hline
TranE                    & 16.54 & 1.3   & 25.86 & 52.50    \\ 
\textbf{TransE-RAA}           & \textbf{17.49} & \textbf{2.1}   & \textbf{26.96} & \textbf{54.55}    \\ \
InsKGC                   & 59.55 & 49.61 & 65.52 & 77.03   \\ 
\textbf{InsKGC-RAA}           & \textbf{62.92} & \textbf{54.31} & \textbf{67.94} & \textbf{78.84}    \\
ComplEx                  & 53.06 & 48.98 & 55.96 & 59.25    \\ 
\textbf{ComplEx-RAA}         & \textbf{55.33} & \textbf{50.96} & \textbf{57.13} & \textbf{62.07}    \\ 
SimKGC                   & 56.23 & 46.63 & 61.83 & 73.83    \\
\textbf{SimKGC-RAA}           & \textbf{63.21} & \textbf{54.47} & \textbf{69.04} & \textbf{77.51}    \\ \hline
\end{tabular}
\caption{Experiments on compatibility of relation-aware anchor enhancement mechanism with other methods on WN18RR\_v1 dataset. The bold numbers indicate the better performance.}
\label{compasion_RAAKGC1}
\end{table}

To assess the compatibility and plug-and-play attribute of our approach, Table \ref{compasion_RAAKGC1} shows the performance of four baseline models w./w.o. our RAA framework on the WN18RR\_v1 dataset \cite{teru2020inductive}. As for triple-based methods, we extend our method to two typical baselines, including TransE \cite{bordes2013translating} and ComplEx \cite{trouillon2017complex}. As for the text-based methods, we choose InsKGC \cite{jiang2023don} and SimKGC \cite{wang-etal-2022-simkgc}. Table \ref{compasion_RAAKGC1} indicates that the RAA framework makes average boosts on all metrics for baselines. Concretely, with the help of our strategy, the performance of InsKGC improves by 3.37\%, 4.7\%, 2.42\%, and 1.81\% on MRR, Hit@1, Hit@3, and Hit@10, respectively. Additionally, SimKGC also shows improvements of 6.98\%, 7.84\%, 7.21\%, and 3.68\% on MRR, Hit@1, Hit@3, and Hit@10. Moreover, ComplEx and TransE also showed improvements on all metrics. These experimental results verify the compatibility of solutions against state-of-the-art algorithms and further suggest that the relation-aware entity information can enhance the discriminative ability of other KGC models.

\subsection{Performance Over Unseen Entities}

\begin{table}
\begin{tabular}{ccccc}
\hline
\multirow{2}{*}{Methods} & \multicolumn{2}{c}{WN18RR\_v1}  & \multicolumn{2}{c}{FB15k-237\_v1}   \\ \cline{2-5} 
                         & MRR            & Hit@10         & MRR            & Hit@10      \\ \hline
\textbf{RAA-KGC}       & \textbf{50.12} & \textbf{84.04} & \textbf{29.11} & \textbf{55.36} \\ 
SimKGC                   & 41.77          & 66.49          & 20.98          & 40.98           \\ 
InsKGC                   & 40.12          & 69.15          & 23.5         & 48.29            \\ \hline   
\end{tabular}
\caption{The Hit@10 and MRR of WN18RR{\_}v1 and FB15k-237{\_}v1 under inductive scenario. The optimal values of each metric are marked in bold.}
\label{compasion_unseen1}
\end{table}

To further explore the generalization capability of RAA-KGC, we conduct experiments under an inductive link prediction setting. The texted datasets include WN18RR\_v1, WN18RR\_v2, WN18RR\_v3, WN18RR\_v4, FB15k-237\_v1, FB15k-237\_v2, FB15k-237\_v3, and FB15k-237\_v4. The eight datasets were extracted by Teru et al. \cite{teru2020inductive}. The compared algorithms include SimKGC and InsKGC \cite{jiang2023don}. Due to space limitations, only the Hit@10 and MRR of WN18RR\_v1 and FB15k-237\_v1 datasets are shown in Table \ref{compasion_unseen1}, more results are presented in the Appendix. As shown in Table \ref{compasion_unseen1}, RAA-KGC outperforms the other two methods on each metric. On the WN18RR\_V1 dataset, RAA-KGC achieves 8.35\% and 10\% higher MRR than SimKGC and InsKGC, respectively. Similarly, on the FB15k-237 dataset, RAA-KGC is 8.13\% and 5.61\% higher than SimKGC and InsKGC on MRR. These results indicate that anchor-enhanced embedding provides more supervision information, enabling the network to better comprehend the semantic information of relations and entities. This allows RAA-KGC to adapt quickly.

\subsection{Case Study}

\begin{table}
\begin{tabular}{ccc}
\hline
\multicolumn{3}{c}{(almaty\_NN\_1,   instance hypernym, urban\_center\_NN\_1)} \\\hline
Model                          & Top5 candidate entites & probabilities  \\\hline
\multirow{5}{*}{SimKGC}        & national\_capital\_NN\_1            & 0.78    \\
                               & dtate\_capital\_NN\_1               & 0.72 \\
                               & capital\_NN\_3                      & 0.65   \\
                               & provincial\_capital\_NN\_1          & 0.57   \\
                               & \textbf{urban\_center\_NN\_1}       & 0.56  \\\hline
\multirow{5}{*}{RAA-KGC}       & \textbf{urban\_center\_NN\_1}       & 0.82    \\
                               & national\_capital\_NN\_1            & 0.68   \\
                               & port\_NN\_1                         & 0.65   \\
                               & capital\_NN\_3                      & 0.57     \\
                               & state\_capital\_NN\_1               & 0.53   \\ \hline 
\end{tabular}
\caption{The ranks predicted and probability by RAA-KGC and SimKGC. Bold indicates the target entity.}
\label{case}
\end{table}

We further intuitively show that the prediction results using \textit{almaty\_NN\_1} as the query head entity and \textit{instance hypernym} as the relation. The top-5 predicted entities and their corresponding probabilities, as ranked by RAA-KGC and SimKGC, are presented in Table \ref{case}. We notice that the RAA-KGC outperforms SimKGC, with the target entity appearing first. The target entity, \textit{urban\_center\_NN\_1}, represents a specific city, while the other predictions denote the broader category of capital. Note that we do not need to know the concrete label of the entity or the relation. What we need to know is that these entities belong to anchors referred to in Def. 1. It is because the relational edges in KGs naturally constrain these entities. Starting from the same query head entity and the same relation, we will reach the tail entities with similar semantics, no matter what the query head entity and relation are. Using such property of the RAA-KGC to construct the relation-aware anchor will be more stable since no other language models are required. 

\section{Conclusion}
In this paper, we propose a relation-aware anchor enhanced knowledge graph completion method, termed RAA-KGC. This method first generates a general example of what might the tail (or head) entity be like, and then pulls its embedding towards the neighborhood of the anchor. Experiments reveal the effectiveness and compatibility of RAA-KGC. In the future, we aim to incorporate the structural and textual information of the subgraph to improve the performance of the KGE model.

\bibliography{aaai25}

\section{Appendix}
In the appendix, we mainly supplement all the results of the two parts of relation-wised performance and performance over unseen entities, which are not fully presented in the main paper due to space constraints. In the relation-wised performance section, to explore which kinds of relations make RAA-KGC excel, we compare the performance of SimKGC and RAA-KGC for each relation on the WN18RR dataset. In the performance over unseen entities section, we show the ability of algorithms to generalize to entirely new entities (unseen entities). And the algorithms include RAA-KGC, SimKGC, and InsKGC in this section. Besides, we also summarize the notations in Table \ref{denote}.


\section{Relation-wised Performance}
\setlength{\parskip}{0pt}
In this section, to further demonstrate which relation improves the performance of the RAA-KGC from a fine-grained perspective, we compare the performance of the algorithms in each relation on WN18RR dataset. We show the details of MRR, Hit@1, Hit@3, and Hit@10. The algorithms include RAA-KGC and SimKGC. These results are shown in Table \ref{comparison_app_relation1}-\ref{comparison_app_relation4}. 

As we can see from these tables, the average value of MRR increased by 2.05\%, Hit@1 by 4.06\%, Hit@3 by 0.33\%, and Hit@10 by 0.09\%. For relation $\_similar\_to$, all indicators reached 100\%, especially the MRR increased from 91.66\% to 100\%. For relations $\_member\_of\_domain\_region$ and $\_member\_of\_domain\_usage$, performance is degraded, and it has been observed that these two relations have a wide range of applications, and they are difficult to locate the type of specific entity. For example, \textit{ben\_NN\_1} and \textit{scottish\_NN\_1} are the neighbors of the query head entity \textit{scotland\_NN\_1} which share the same relation \textit{\_member\_of\_domain\_region}, but \textit{ben\_NN\_1} and \textit{scottish\_NN\_1} have completely different meanings, where \textit{ben\_NN\_1} means \textit{a mountain or tall hill} and \textit{scottish\_NN\_1} means \textit{the dialect of English used in Scotland}.

\begin{table}[b]
\centering
\small
\scalebox{1.0}{
\begin{tabular}{@{}ll@{}}
\toprule
\textbf{Notation} & \textbf{Explanation}  \\ \midrule
$\mathcal{G}$  & Knowledge graph (KG)    \\
$\mathcal{E}$ & Entitiy set in KG  \\
$\mathcal{R}$    & Relation set in KG    \\
$\mathcal{G}_{inv}$  & Knowledge graph with all the edges inversed  \\
$\mathcal{T}$ & Relation-aware entity set \\
$\mathcal{T}_k$ & Anchor set \\
$\mathcal{D}_i$   & Anchor-enhanced queries \\
$g_{_1}(\cdot)$ & First knowledge graph encoder \\
$g_{_2}(\cdot)$ & Second knowledge graph encoder \\
$\bm{e_{hr}}$  & Relation-aware embedding\\
$\bm{e_{hrt_{a}}}$ & Anchor-enhanced queries embedding \\
$\bm{e_{hrt_{a}}^{avg}}$  & Anchor-enhanced embedding generation\\
$\bm{e_{t}}$  & Candidate embedding\\ \bottomrule
\end{tabular}}
\caption{Notation summary.}
\label{denote} 
\end{table}

\begin{table}
\centering
\resizebox{0.8\linewidth}{!}{
\begin{tabular}{ccccccccccccc}
\hline
Relations                         & SimKGC         & RAA-KGC   \\ \hline
\_also\_see                       & 63.99          & \textbf{67.69}  \\ 
\_derivationally\_related\_form & 86.12          & \textbf{87.4}   \\ 
\_has\_part                       & 41.08          & \textbf{43.66}  \\ 
\_hypernym                        & 40.17          & \textbf{41.72} \\ 
\_Instance\_hypernym              & 65.56          & \textbf{68.38} \\ \
\_member\_meronym                 & 43.09          & \textbf{44.85}  \\ 
\_member\_of\_domain\_region      & \textbf{45.38} & 41.89      \\ 
\_member\_of\_domain\_usage       & \textbf{47.13} & 44.95    \\ 
\_similar\_to                   & 91.66          & \textbf{100}   \\ 
\_synset\_domain\_topic\_of       & 55.01          & \textbf{57.95}  \\ 
\_verb\_group                     & 91.03          & \textbf{94.27}  \\ \hline
avg                               & 60.92          & \textbf{62.97}  \\ \hline
\end{tabular}
}
\caption{MRR of each relation on WN18RR dataset. The \textbf{boldface} values indicate the best results.}
\label{comparison_app_relation1}
\end{table}

\begin{table}
\centering
\resizebox{0.8\linewidth}{!}{
\begin{tabular}{ccccccccccccc}
\hline
Relations                         & \multicolumn{1}{c}{SimKGC} & \multicolumn{1}{c}{RAA-KGC}  \\ \hline
\_also\_see                       & 58.04                      & \textbf{65.18}                \\ 
\_derivationally\_related\_form & 77.93                      & \textbf{79.84}                \\ 
\_has\_part                       & 29.94                      & \textbf{33.14}                 \\ 
\_hypernym                        & 30.25                      & \textbf{32.37}                 \\ 
\_Instance\_hypernym              & 53.28                      & \textbf{58.6}                  \\ 
\_member\_meronym                 & 33.59                      & \textbf{35.38}                  \\ 
\_member\_of\_domain\_region      & \textbf{38.46}             & 34.62                          \\ 
\_member\_of\_domain\_usage       & \textbf{37.5}              & \textbf{37.5}                  \\ 
\_similar\_to                   & 83.34                      & \textbf{100}                  \\ 
\_synset\_domain\_topic\_of       & 47.37                      & \textbf{51.31}           \\ 
\_verb\_group                     & 83.34                      & \textbf{89.74}                 \\ \hline
avg                               & 52.09                      & \textbf{56.15}                \\ \hline
\end{tabular}
}
\caption{Hit@1 of each relation on WN18RR dataset. The \textbf{boldface} values indicate the best results.}
\label{comparison_app_relation2} 
\end{table}

\begin{table}
\centering
\resizebox{0.8\linewidth}{!}{
\begin{tabular}{ccccccccccccc}
\hline
Relations                         & \multicolumn{1}{c}{SimKGC} & \multicolumn{1}{c}{RAA-KGC}  \\ \hline
\_also\_see                       & 67.85                      & \textbf{67.86}                \\ 
\_derivationally\_related\_form & \textbf{94.04}                      & \textbf{94.04}                 \\ 
\_has\_part                       & 46.22                      & \textbf{47.68}              \\
\_hypernym                        & 44.41                      & \textbf{45.56}               \\ 
\_Instance\_hypernym              & 74.17                      & \textbf{76.22}                 \\ 
\_member\_meronym                 & 46.25                      & \textbf{49.01}                 \\ 
\_member\_of\_domain\_region      & \textbf{48.08}             & 46.15                        \\ 
\_member\_of\_domain\_usage       & \textbf{52.08}             & 47.92                        \\ 
\_similar\_to                   & \textbf{100}                        & \textbf{100}                \\ 
\_synset\_domain\_topic\_of       & 57.9                       & \textbf{61.4}             \\ 
\_verb\_group                     & \textbf{100}               & 98.72                          \\ \hline
avg                               & 66.45                      & \textbf{66.78}                \\ \hline
\end{tabular} 
}
\caption{Hit@3 of each relation on WN18RR dataset. The \textbf{boldface} values indicate the best results.}
\label{comparison_app_relation3}
\end{table}

\begin{table}
\centering
\resizebox{0.9\linewidth}{!}{
\begin{tabular}{ccccccccccccc}
\hline
Relations                         & \multicolumn{1}{c}{SimKGC} & \multicolumn{1}{c}{RAA-KGC} & \\ \hline
\_also\_see                       & 74.11                      & \textbf{76.22}                 \\ 
\_derivationally\_related\_form & 98.19                      & \textbf{98.32}                 \\ 
\_has\_part                       & 63.95                      & \textbf{65.99}                 \\ 
\_hypernym                        & 59.16                      & \textbf{59.63}                 \\ 
\_Instance\_hypernym              & 86.02                      & \textbf{87.83}                  \\ 
\_member\_meronym                 & 62.45                      & \textbf{63.04}                \\ 
\_member\_of\_domain\_region      & \textbf{55.77}             & \textbf{55.77}                 \\
\_member\_of\_domain\_usage       & \textbf{68.75}             & 62.5                            \\ 
\_similar\_to                   & \textbf{100}               & \textbf{100}                 \\ 
\_synset\_domain\_topic\_of       & 68.86                      & \textbf{71.93}                \\ 
\_verb\_group                     & \textbf{100}               & \textbf{100}                  \\ \hline
avg                               & 76.11                      & \textbf{76.20}             \\ \hline
\end{tabular}
} 
\caption{Hit@10 of each relation on WN18RR dataset. The \textbf{boldface} values indicate the best results.}
\label{comparison_app_relation4} 
\end{table}

\vspace{100pt}
\section{Performance over Unseen Entities}

RAA-KGC shows much greater ability to generalize to entirely new entities (unseen entities) that are added to the knowledge graph as evidenced by the much superior performance compared to baselines. We hypothesize that this is due to the power of RAA-KGC to generalize to unseen nodes added to a graph as well as entirely new graphs. We show the results of MRR, Hit@1, Hit@3, Hit@10 on the inductive link prediction setting. The texted datasets include WN18RR\_v1, WN18RR\_v2, WN18RR\_v3, WN18RR\_v4, FB15k-237\_v1, FB15k-237\_v2, FB15k-237\_v3, and FB15k-237\_v4. The compared algorithms include SimKGC and InsKGC. These results are shown in Table \ref{comparison_app_unseen1}-\ref{comparison_app_unseen8}. 

As we can see from these tables, in inductive scenario, the performance improvement is more significant, because RAA-KGC indicates that learning anchor-enhanced embedding provides more supervision information and drives the network to achieve a better understanding of the semantic information of relations and entities. This allows RAA-KGC to adapt quickly. For example, on the WN18RR\_v1 dataset, the MRR index of RAA-KGC is 8.35\% higher than that of SimKGC and 10\% higher than that of InsKGC. Similarly, other datasets are similar.
\begin{table}[h]
\begin{tabular}{ccccc}
\hline
\multirow{2}{*}{Method} & \multicolumn{4}{c}{WN18RR\_v1}  \\ \cline{2-5} 
                        & MRR            & Hit@1          & Hit@3          & Hit@10         \\ \hline
RAA-KGC               & \textbf{50.12} & \textbf{33.25} & \textbf{61.97} & \textbf{84.04}  \\ 
SimKGC                  & 41.77          & 28.73          & 47.61          & 66.49           \\ 
InsKGC                  & 40.12          & 25.26          & 48.14          & 69.15          \\ \hline
\end{tabular}
\caption{The results in WN18RR{\_}v1 dataset. The \textbf{boldface} values indicate the best results.}
\label{comparison_app_unseen1}  
\end{table}

\begin{table}[h]
\begin{tabular}{ccccc}
\hline
\multirow{2}{*}{Method} & \multicolumn{4}{c}{WN18RR\_v2} \\ \cline{2-5} 
                        & MRR                                & Hit@1                              & Hit@3                              & Hit@10   \\ \hline
RAA-KGC              & \multicolumn{1}{l}{\textbf{55.74}} & \multicolumn{1}{l}{\textbf{41.38}} & \multicolumn{1}{l}{\textbf{63.71}} & \multicolumn{1}{l}{\textbf{85.15}}  \\ 
SimKGC                  & \multicolumn{1}{l}{28.63}          & \multicolumn{1}{l}{11.34}          & \multicolumn{1}{l}{39.01}          & \multicolumn{1}{l}{60.55}           \\ 
InsKGC                  & \multicolumn{1}{l}{45.73}          & \multicolumn{1}{l}{34.02}          & \multicolumn{1}{l}{34.02}          & \multicolumn{1}{l}{69.5}         \\ \hline
\end{tabular}
\caption{The results of WN18RR{\_}v2 dataset. The \textbf{boldface} values indicate the best results.}
\label{comparison_app_unseen2}  
\end{table}

\begin{table}[h]
\begin{tabular}{ccccc}
\hline
\multirow{2}{*}{Method} & \multicolumn{4}{c}{WN18RR\_v3}  \\ \cline{2-5}
                        & MRR            & Hit@1          & Hit@3         & Hit@10         \\ \hline
RAA-KGC              & \textbf{32.41} & \textbf{23.06} & \textbf{37.7} & \textbf{54.15}   \\
SimKGC                  & 27.57          & 17.94          & 30.74         & 47.6           \\
InsKGC                  & 32.1           & 22.48          & 36.78         & 50.16         \\ \hline
\end{tabular}
\caption{The results of WN18RR{\_}v3 dataset. The \textbf{boldface} values indicate the best results.}
\label{comparison_app_unseen3}  
\end{table}

\begin{table}[H]
\begin{tabular}{ccccc}
\hline
\multirow{2}{*}{Method} & \multicolumn{4}{c}{WN18RR\_v4}  \\ \cline{2-5} 
                        & MRR   & Hit@1 & Hit@3 & Hit@10    \\ \hline
RAA-KGC               & \textbf{50.26} & \textbf{35.97} & \textbf{59.79} & \textbf{76.31}   \\ 
SimKGC                  & 31.33 & 15.82 & 39.54 & 61.58   \\
InsKGC                  & 38.64 & 27.22 & 44.58 & 59.97   \\ \hline
\end{tabular}
\caption{The results of WN18RR{\_}v4 dataset. The \textbf{boldface} values indicate the best results.}
\label{comparison_app_unseen4}  
\end{table}

\begin{table}[H]
\begin{tabular}{ccccc}
\hline
\multirow{2}{*}{Method} & \multicolumn{4}{c}{FB15k-237\_v1}  \\ \cline{2-5} 
                        & MRR            & Hit@1          & Hit@3          & Hit@10       \\ \hline
RAA-KGC               & \textbf{29.11} & \textbf{16.58} & \textbf{35.37} & \textbf{55.36}  \\ 
SimKGC                  & 20.98          & 10             & 25.37          & 40.98         \\ 
InsKGC                  & 23.5           & 11.95          & 27.32          & 48.29        \\ \hline
\end{tabular}
\caption{The results of FB15k-237{\_}v1 dataset. The \textbf{boldface} values indicate the best results.}
\label{comparison_app_unseen5}  
\end{table}

\begin{table}[H]
\begin{tabular}{ccccc}
\hline
\multirow{2}{*}{Method} & \multicolumn{4}{c}{FB15k-237\_v2} \\ \cline{2-5} 
                        & MRR            & Hit@1          & Hit@3          & Hit@10       \\ \hline
RAA-KGC               & \textbf{33.64} & \textbf{21.76} & \textbf{37.97} & \textbf{59.41}  \\ 
SimKGC                  & 21.77          & 11.5           & 24.79          & 41.84            \\ 
InsKGC                  & 22.13          & 11.72          & 24.58          & 44.25         \\ \hline
\end{tabular}
\caption{The results of FB15k-237{\_}v2 dataset. The \textbf{boldface} values indicate the best results.}
\label{comparison_app_unseen6}  
\end{table}

\begin{table}[H]
\begin{tabular}{ccccc}
\hline
\multirow{2}{*}{Method} & \multicolumn{4}{c}{FB15k-237\_v3}  \\ \cline{2-5} 
                        & MRR            & Hit@1         & Hit@3          & Hit@10      \\ \hline
RAA-KGC               & \textbf{26.63} & \textbf{16.3} & \textbf{30.81} & \textbf{46.3}   \\ 
SimKGC                  & 18.92          & 9.94          & 20.57          & 36.24        \\ 
InsKGC                  & 18.22          & 8.62          & 19.77          & 38.73         \\ \hline
\end{tabular}
\caption{The results of FB15k-237{\_}v3 dataset. The \textbf{boldface} values indicate the best results.}
\label{comparison_app_unseen7}  
\end{table}

\begin{table}[H]
\begin{tabular}{ccccc}
\hline
\multirow{2}{*}{Method} & \multicolumn{4}{c}{FB15k-237\_v4}  \\ \cline{2-5} 
                        & MRR            & Hit@1          & Hit@3          & Hit@10      \\ \hline
RAA-KGC               & \textbf{26.49} & \textbf{16.19} & \textbf{29.56} & \textbf{48.7} \\ 
SimKGC                  & 19.82          & 11.3           & 20.92          & 37.64        \\ 
InsKGC                  & 20.01          & 10.36          & 23.28          & 39.92         \\ \hline
\end{tabular}
\caption{The results of FB15k-237{\_}v4 dataset. The \textbf{boldface} values indicate the best results.}
\label{comparison_app_unseen8}  
\end{table}

\end{document}